\documentclass[11pt,twoside]{article}
\usepackage{asp2010}

\resetcounters

\begin{document}

\title{AGB Stars in WLM}
\author{Benjamin Tatton$^1$, Maria-Rosa Cioni$^1$, and Mike Irwin$^2$
\affil{$^1$Centre for Astrophysics Research, Science \& Technology Research Institute, University of Hertfordshire, Hatfield, AL10 9AB, U.K.}
\affil{$^2$Cambridge Astronomical Survey Unit, Institute of Astronomy,\\ Madingley Road, Cambridge CB4 0HA, U.K.}}

\begin{abstract}
We investigate the star formation history and metallicity of the Local Group irregular dwarf galaxy WLM using wide-field JHK near-infrared imaging, spanning a region of approximately 1 sq. degree, obtained with WFCAM on UKIRT. JHK photometry clearly reveals the tip of the red giant branch, allowing a new estimate of the distance, and allows ready identification of C-type and M-type AGB stars. The C/M ratio was used to produce a surface map of the metallicity distribution which is compared to previous studies.  Multi-wavelength spectral energy distributions (SEDs) were constructed for some AGB stars.
\end{abstract}

\section{Introduction and Data}
WLM (Wolf-Lundmark-Melotte) is a dwarf irregular galaxy and is a member of the Local Group. It is at a distance of $932\pm33$ kpc \citep{mcconnachie05}. In \citeyear{valcheva07} \citeauthor{valcheva07} did a photometric study on part of WLM and found a C/M ratio of $0.56\pm0.12$ which differed greatly (by a factor of 20) from previous values. In \citeyear{leaman09} \citeauthor{leaman09} used Ca-II triplet spectroscopy on $78$ red giant stars and found a mean [Fe/H] of $-1.27\pm0.04$ dex and also found that stars closer to the centre of the galaxy were more metal rich by $0.30\pm0.06$ dex.
The data used here is near-infrared (NIR) $JHK$ observations made on October $16^\mathrm{th}$ $2007$ using WFCAM on UKIRT in Hawaii.

\section{Results}

\subsection{C/M ratio and Metallicity}
The C/M ratio represents the number ratio between carbon rich (C-type) and oxygen-rich (M-type) asymptotic giant branch (AGB) stars. Using a histogram of $J-K$ colour we find a C to M split at $J-K=1.05\pm0.05$ mag which gives ratios of $0.27$ to $0.89$ for the inner half square degree of data depending on foreground removal methods. In the central area dominated by the galaxy (an ellipse with RA=$\pm0.07^{\circ}$ Dec$=\pm0.15^{\circ}$) we obtain ratios between $0.4$ to $0.8$. When we correct our C/M ratio using the C-type catalogue of \citet{battinelli03} (to account for flaws in using $J-K$ colour as cut-off) these ratios range from $0.36$ to $1.43$ and $0.55$ to $1.24$ for inner and central regions respectively. The lower ratio of the central region compared to the inner region could be due to a metal rich star-forming region within the central part of WLM. Using the same sky area as \citet{valcheva07}, we on the whole find our reduced data agrees more with their unreduced data and vice versa which could be due to their adopted foreground having a large number of C-type stars. 
The C/M ratio is calibrated to [Fe/H] using the equation B.1 from \citet{cioni09}. By applying this equation to our data we obtain [Fe/H] values from $-1.12$ to $-1.37$ dex for original C/M ratios and $-1.18$ to $-1.43$ dex for corrected ones for the inner field. For the central field these values are from $-1.20$ to $-1.34$ dex and $-1.27$ to $-1.40$ dex for original and corrected C/M ratios.

\subsection{Distance modulus (m-M)}
The tip of the RGB (TRGB) represents the split between the RGB and AGB populations. The TRGB is found at $K=18.7\pm0.1$ mag and is used to calculate the distance modulus once a value for its absolute magnitude is known. We explored two methods of obtaining this value; evolutionary tracks and [Fe/H]. With the tracks by \citet{marigo08} we did two calculations, one for age ranges and the other for metallicity giving distance moduli of (m-M)$=24.37$ mag at a constant metallicity and varying age, and (m-M)$=24.39$ mag at a constant age and varying metallicity. For the calculated [Fe/H] we make use of the relation between [Fe/H] and the absolute magnitudes of RGB stars, by \citet{ferraro00}.
Here, it is assumed the overall metallicity of the galaxy is the same for both the AGB and RGB population. The mean of the values for the different methods explored is (m-M)$=24.89\pm0.25$ mag ($\sim951$ kpc). This value agrees with previous measurements.

\subsection{Spectral Energy Distribution (SED)}
By combining our NIR data with optical data from \citet{mcconnachie05} and mid-infrared data from \citet{boyer09} we can investigate the SED of AGB stars in WLM. The SED allows us to obtain bolometric fluxes (and bolometric magnitudes from the distance modulus). We found $1281$ matches between all the datasets after applying some small systematic shifts. For the sources to be usable in a SED there needed to be a magnitude present in every band, in total $52$ sources met this criteria. When deriving the bolometric flux we found that the most luminous stars were not the C-type AGBs but supergiants. The bolometric fluxes were also converted into bolometric magnitudes giving us additional data to confirm stellar types.

\bibliography{tatton}

\end{document}